\documentclass[journal]{IEEEtran}
\usepackage{latexsym,epsf,times,amsmath,color,amssymb,indentfirst,fancyhdr,colortbl,
bm}
\usepackage{subcaption}
 \usepackage{graphicx}
\newcommand{\nc}{\newcommand}
\nc{\be}{\begin{equation}}
\nc{\ee}{\end{equation}}

\def\calH{{\mathcal{H}}}

\def\setI{{\mathbb{I}}}
\def\setZ{{\mathbb{Z}}}

\nc{\Abf}{\mathbf{A}}
\nc{\Cbf}{\mathbf{C}}
\nc{\Dbf}{\mathbf{D}}
\nc{\Fbf}{\mathbf{F}}
\nc{\Gbf}{\mathbf{G}}
\nc{\Hbf}{\mathbf{H}}
\nc{\Ibf}{\mathbf{I}}
\nc{\Nbf}{\mathbf{N}}
\nc{\Qbf}{\mathbf{Q}}
\nc{\Rbf}{\mathbf{R}}
\nc{\Sbf}{\mathbf{S}}
\nc{\Ubf}{\mathbf{U}}
\nc{\Xbf}{\mathbf{X}}
\nc{\Ybf}{\mathbf{Y}}
\nc{\Zbf}{\mathbf{Z}}
\nc{\abf}{\mathbf{a}}
\nc{\bbf}{\mathbf{b}}
\nc{\hbf}{\mathbf{h}}
\nc{\gbf}{\mathbf{g}}
\nc{\nbf}{\mathbf{n}}
\nc{\pbf}{\mathbf{p}}
\nc{\qbf}{\mathbf{q}}
\nc{\rbf}{\mathbf{r}}
\nc{\sbf}{\mathbf{s}}
\nc{\ubf}{\mathbf{u}}
\nc{\vbf}{\mathbf{v}}
\nc{\xbf}{\mathbf{x}}
\nc{\ybf}{\mathbf{y}}
\nc{\zbf}{\mathbf{z}}
\nc{\unobf}{\mathbf{1}}
\nc{\zerobf}{\mathbf{0}}
\nc{\vbfs}{\mathbf{\scriptsize v}}
\nc{\Phibf}{\mathbf{\Phi}}
\nc{\Psibf}{\mathbf{\Psi}}
\nc{\Lambdabf}{\mathbf{\Lambda}}
\nc{\Sigmabf}{\mathbf{\Sigma}}
\nc{\Omegabf}{\mathbf{\Omega}}
\nc{\xibf}{{\mbox{\boldmath $\xi$}}}
\nc{\xibfs}{{\mbox{\boldmath \scriptsize $\xi$}}}
\nc{\diag}{\text{diag}}
\nc{\sign}{\text{sign}}
\nc{\E}{\text{E}}
\def\defeq{\stackrel {\scriptscriptstyle \Delta}{=}}

\IEEEoverridecommandlockouts
\makeatletter
\title{Pseudo-Zernike Moments Based Sparse Representations for SAR Image Classification}
\author{\IEEEauthorblockN{Shahzad Gishkori and Bernard Mulgrew}
\thanks{S. Gishkori and B. Mulgrew are with Institute for Digital Communications (IDCOM), The School of Engineering, The University of Edinburgh, UK. 
Emails: \{s.gishkori, b.mulgrew\}@ed.ac.uk}
\thanks{This work was supported by Jaguar Land Rover and the UK-EPSRC grant EP/N012240/1 as part of the jointly funded Towards Autonomy: Smart and Connected Control (TASCC) Programme.}
}
\providecommand{\keywords}[1]{\textbf{\textit{Index terms---}} #1}
\begin{document}
\maketitle
\vspace{-0.3in}
\begin{abstract}
\noindent
We propose radar image classification via pseudo-Zernike moments based sparse representations. 
We exploit invariance properties of pseudo-Zernike moments to augment redundancy in the sparsity representative dictionary by introducing auxiliary atoms. 
We employ complex radar signatures.
We prove the validity of our proposed methods on the publicly available MSTAR dataset.
\end{abstract}
\section{Introduction}
\label{sect:intro}
Synthetic aperture radar (SAR) can provide all-weather imagery with a very high resolution \cite{Carrara_95}. This has naturally led to using SAR for the purpose of automatic target recognition or classification. Initial usage was military related. However, SAR imaging with the aim of classification is making very quick strides for the automotive usage as well \cite{SAR_Mobile_Mapping_16}. 
Traditionally, a number of techniques are used for SAR image classification. Here, we briefly mention a couple of them.
Template based classification \cite{Template-based_Classify_99} requires generation of a large number of templates for each target and then matching the test image with  those templates in an exhaustive search manner. It is an effective linear approach. However, it is computationally quite expensive. 
Among the nonlinear approaches, support vector machine classifier (SVC) has been quite popular \cite{zhao_sar_svm_01}. It is a large margin classifier and it can outperform the template based classifier. However, this approach is dependent upon accurate estimation of the pose angle which involves an extra preprocessing stage.
\\
Recent trends in classification are based on sparse representations, also known as sparse coding \cite{olshausen96,BP}. Initially, efforts were made to find or use a unified dictionary for all the classes, see, e.g., \cite{ksvd_06,Huang_SRC_06} and references therein. Instead of using a single dictionary for all the classes, \cite{src_09} proposed to use unit normalised measurements of the objects as the columns of an overcomplete dictionary. Coding is done through an $\ell_1$-norm minimisation problem and the classification is based on a least-squares metric w.r.t. the group of columns specific to a particular class object. This is known as sparse representation based classifier (SRC). The ease of formulating a dictionary by using the measurements of the class objects directly, made SRC a favourable choice for classification in a wide range of fields. In SAR image classification, SRC was used in \cite{src_radar_10} from the perspective of class manifolds \cite{manifold_05}. A class manifold is defined over the set of measurements for a particular class object, and the SAR image is claimed to lie in that manifold by using the fact that linear representation can be provided to a nonlinear manifold if a local region of the manifold is considered \cite{Roweis_00}. This local region of the manifold gives the basis for sparse representation of a test image over that manifold. Therefore, it obviates the need for a rigorous preprocessing as well as pose angle estimation.
Dimensionality reduction can be achieved via random projections. However, it can result in a performance loss.
\\
It was shown in \cite{src_09,src_radar_10} that SRC can outperform linear SVC (LSVC), i.e., when a linear kernel is used. However, SRC is primarily based upon sparse reconstruction or coding, and it does not involve the classification aspect during the coding process. A number of papers have been written to incorporate this aspect in sparse representations. Some discriminative dictionary learning techniques have been proposed in \cite{Yang_metaface_10,Ramirez_10}. Similarly, joint dictionary learning and encoding has been proposed in \cite{Yang_fisher_11}. Although, these methods provide good performance but dictionary learning, whether discriminative or not, is a computationally intensive process.
\\
Moments based image representations have been successfully used over many decades \cite{Hu_moments_62,Sadjadi_3Dmoments_1980}. 
The basic idea is to derive image features which are scale-, shift- and rotation-invariant by using nonlinear combinations of the regular moments (also called geometric moments). However, the gains have been limited, primarily due to the non-orthogonality of regular moments. Orthogonal moments, e.g., Legendre moments, Zernike moments and pseudo-Zernike (PZ) moments \cite{Teh_moments_88,zern_moments_90} have been a popular substitute in pattern recognition. Among these, PZ-moments stand apart both in terms of generating the maximum number of invariant moments as well as in terms of performance regarding noise rejection. PZ-moments have been used for radar automatic target recognition in \cite{Clemente_PZ_14} with a nearest neighbour classifier. 
Similarly, in \cite{Clemente_PZ_15}, PZ-moments have been used for radar classification based on its micro-Doppler signatures, with an SVC. However, in both these cases, the emphasis has been on feature extraction w.r.t. PZ-moments and not on the choice of an optimal classifier.
\\
\noindent
{\bf \em Contributions}.
In this paper, we propose using PZ-moments in combination with the SRC framework (PZ-SRC), in order to gain from both optimal feature extraction as well as optimal classification. 
By using a finite number of PZ-moments, we reduce the dimensionality of the problem.
Due to invariance properties of the PZ-moments, we obtain good performance, albeit in the low dimensional setting. 
We also introduce auxiliary atoms in the dictionary to increase the redundancy of information, which further exploits the invariance properties of the PZ-moments. Thus, information is better localised in individual class manifolds. This results in a further improvement in classification performance.
Note, this also forms a unique contribution to the SRC framework, in general, as well.
In order to utilise both the magnitude as well as the phase information of the complex radar signatures, we fuse the two parameters by a simple averaging mechanism (see \cite{Sadjadi_fusion_05} for details on the fusion mechanisms). This results in an even more informative radar signatures with direct positive impact over the classification performance. 
Note, a similar approach has been used in \cite{Sadjadi_SRC_16}. However, the feature extraction there is based on regular moments and the authors do not use auxiliary atoms.
Now, in order to encode the test image in our proposed framework of PZ-SRC, we use state-of-the-art technique of iterative hard thresholding (IHT) algorithm \cite{Davies_IHT}. IHT provides very fast convergence as well accuracy (in comparison to the approach of \cite{src_09}), both of which are crucial in real-time radar image classification applications. We test our proposed methods on the publicly available MSTAR dataset.
\\
\noindent
{\bf \em Organisation}.
Section 
\ref{sec:pz_moments} gives the basics of PZ-moments, Section \ref{sec:src} briefly describes the SRC method, Section \ref{sec:prop} details our proposed method of PZ-SRC,  Section \ref{sec:sim} provides simulation results and conclusions are given in Section \ref{sec:concl}.
\\
\noindent
{\bf \em Notations}.
Matrices are in upper case bold while column vectors are in lower case bold,
$(\cdot)^T$ denotes transpose,
$[\xbf]_{i}$ is the $i$th element of $\xbf$,
$\hat{\xbf}$ is the estimate of $\xbf$,
$\defeq$ defines an entity,
and 
the $\ell_p$-norm is denoted as $||\xbf||_p = (\sum_{i=0}^{N-1} |[\xbf]_{i}|
^p)^{1/p}$.
\section{Pseudo-Zernike Moments}
\label{sec:pz_moments}

Let a piecewise continuous function $s(x,y)$ (with bounded support) be the intensity function of  a $2$-D real image in Cartesian coordinates.
The regular moments of $s(x,y)$ can be defined as
\be
\mu_{p,q} = \int_x \int_y x^p y^q s(x,y) \, dx\, dy
\label{eq:reg_mom}
\ee
where $\{p,q\} \in \setZ_+$ and $p+q$ is the degree of the moments. Note, (\ref{eq:reg_mom}) represents the projection of $s(x,y)$ on monomial $x^p y^q$.  Since $\{x^p y^q\} $ is not an orthogonal set, $\mu_{p,q}$ are not independent moments. 
In contrast, the PZ-moments are generated from a set of orthogonal polynomials. We refer to these polynomials as PZ-polynomials.
The PZ-polynomials are a set of complex polynomials described as
\be
z_n^m (r,\theta) = \rho_n^m(r) \exp \left(j m \theta \right)
\label{eq:pz_poly}
\ee
where $r \defeq \sqrt{x^2+y^2}$ and $\theta \defeq \tan^{-1}(y/x)$ are the length and angle of the position vector of a point $(x,y)$ w.r.t. the centre of the image, respectively, $n\in \setZ_+$ is the degree of the polynomial with frequency $m$, i.e., $m \in [-n,+n]$, and
\be
\rho_n^m(r) \defeq \sum_{\kappa}^{n-|m|} \dfrac{ (-1)^\kappa \, (2n+1-\kappa)! \, r^{n-\kappa}} 
{\kappa! \, (n+|m|+1-\kappa)! \, (n-|m|-\kappa)!}
\label{eq:rad_poly}
\ee
is the radial polynomial. When defined over a unite circle, i.e., $r\le 1$, the PZ-polynomials exhibit orthogonality, i.e., 
\be
\int_0^{2\pi} \int_0^1 [z_n^m (r,\theta) ]^* z_{n'}^{m'} (r,\theta)\, r \, dr \, d\theta = \dfrac{\pi}{n+1}\delta_{nn'}\delta_{mm'}
\label{eq:pz_orth}
\ee
where $\delta_{ii'}$ is the Kronecker delta function. 
Note, it can be seen via simple enumeration that cardinality of the set of PZ-polynomials with $\text{degree}\le n$ is, $P = (n+1)^2$.
Now, the PZ-moments can be obtained by projecting the image onto the PZ-polynomials as\footnote{Note, in case of a digital image, the integrals in the projection operations are replaced by summations.}
\be
a_n^m = \dfrac{n+1}{\pi} \int_0^{2\pi} \int_0^1 [z_n^m (r,\theta) ]^* s(r,\theta)\, r \, dr \, d\theta
\label{eq:pz_mom}
\ee
where $s(r,\theta) = s(x,y)|_{x=r\cos\theta,y=r\sin\theta}$.
Due to (\ref{eq:pz_orth}), it can be shown that (\ref{eq:pz_mom}) generates a set of independent moments.
\\
The invariance properties of the PZ-moments can be established via mathematical manipulations. For scale and translation invariance, one way is to use the regular moments of the image. The transformed image can be written as
\be
g(x,y) = s\left(\frac{x}{\upsilon}+m_x, \frac{y}{\upsilon}+ m_y\right)
\label{eq:invar_scaleTr}
\ee
where $m_x\defeq \mu_{1,0}/\mu_{0,0}$ and $m_y\defeq\mu_{0,1}/\mu_{0,0}$ are the centroid adjustment parameters of the image $s(x,y)$, and $\upsilon \defeq \sqrt{\xi/\mu_{0,0}}$ is the scale adjustment parameter of the image $s(x,y)$ with a predetermined value $\xi$. Now, the scale- and translation-invariant PZ-moments can be generated by replacing $s(r,\theta)$ with $g(r,\theta)$ in (\ref{eq:pz_mom}), where $g(r,\theta) = g(x,y)|_{x=r\cos\theta,y=r\sin\theta}$.
Since PZ-polynomials are a set of complex polynomials, the PZ-moments generated via (\ref{eq:pz_mom}) are also complex. The rotation invariance of the PZ-moments refers to the magnitude part only, i.e., $|a_n^m|$ and not the phase.
\section{Sparse Representation Based Classifier}
\label{sec:src}
Let a generic $\sqrt{N}\times\sqrt{N}$ image with intensity function $g(x,y)$ (or $g(r,\theta)$) is represented as an $N\times 1$ vector $\gbf$ via a lexicographic ordering (column or row ordered). 
Let $\gbf_j^k$ be the $j$th image measurement of the $k$th object class, for $j=1,\cdots,J_k$ and $k=1,\cdots,K$. Now, given a set of training image measurements $\{\gbf_j^k\}$, with $\|\gbf_j^k\|_2^2=1$, the SRC method defines the dictionary as
\be
\Gbf \defeq [ \Gbf^1, \Gbf^2, \cdots, \Gbf^K]
\label{eq:G_dict}
\ee
where $\Gbf$ is an $N\times J$ matrix with $J = \sum_{k=1}^K J_k$ and $\Gbf^k \defeq [\gbf_1^k, \gbf_2^k, \cdots, \gbf_{J_k}^k]$ is an $N\times J_k$ matrix acting as a sub-dictionary for class $k$, for $k=1,\cdots,K$.
Any test image measurement, represented as an $N \times 1$ vector $\tilde{\ybf}$ can then be decomposed or encoded according to the linear model
\be
\tilde{\ybf} = \Gbf \tilde{\xbf} + \tilde{\nbf}
\label{eq:src_model}
\ee
where $\tilde{\xbf}$ is a $J\times 1$ vector of coefficients defined as, 
$\tilde{\xbf} \defeq [ \tilde{\xbf}^{1\,T}, \tilde{\xbf}^{2\,T}, \cdots, \tilde{\xbf}^{K\,T}]^T$,
where $\tilde{\xbf}^{k}$ are the coefficients w.r.t. the sub-matrix $\Gbf^k $, and the $N \times 1$ vector $\tilde{\nbf}$ accounts for model errors with a bounded energy, i.e., $\| \tilde{\nbf}\|_2<\tilde{\epsilon}$. It is clear from (\ref{eq:src_model})
that given $\tilde{\ybf}$ belongs to the $k$th class, $\tilde{\xbf}$ would be a sparse vector. Now, an estimate of $\tilde{\xbf} $ can be obtained by solving the following $\ell_1$-norm optimisation problem (OP).
\be
\hat{\tilde{\xbf}} =  \arg \min_{\tilde{\xbf}}  \left\| \tilde{\ybf} - \Gbf \tilde{\xbf}\right\|_2^2 + \lambda \left\| \tilde{\xbf}\right\|_1^1
\label{eq:l1}
\ee
where $\lambda>0$. The classification result is then obtained by finding the $k$ for which 
$\| \tilde{\ybf} - \Gbf^k \hat{\tilde{\xbf}}^k \|_2^2$ is minimum, for $k=1, \cdots, K$.
In case of feature based representation, the SRC model takes the form,
$\Rbf \tilde{\ybf} = \Rbf \Gbf \tilde{\xbf} + \Rbf  \tilde{\nbf}
$,
where $\Rbf$ is an $R\times N$ linear transformation matrix. Generally, $\Rbf$ is a random matrix with $R\ll N$.
\section{PZ-moments based Sparse Representations}
\label{sec:prop}
In this paper we consider feature based sparse representations. In our case, PZ-moments form the feature set of the radar image. Since PZ-moments are generated by projecting the image onto PZ-polynomials, we can easily generate the features by converting the PZ-polynomials into a basis matrix and then projecting the image vector onto this basis matrix. 
\\
Let a $P\times 1$ vector $\zbf_i$ of PZ-polynomials, with $\text{degree}\le n$, w.r.t. image point $(r_i,\theta_i)$, where $(r_i,\theta_i)$ are the Polar coordinates equivalent of the image point $(x_i,y_i)$ in Cartesian coordinates, for $i = 1,\cdots,N$, be defined as
\begin{align}
\zbf_i &\defeq [\gamma_0 z_0^0 (r_i,\theta_i),
\gamma_1 z_1^{-1} (r_i,\theta_i), \gamma_1 z_1^{0} (r_i,\theta_i), \gamma_1 z_1^{+1} (r_i,\theta_i), \notag \\
& \cdots,  \gamma_n z_n^{-n} (r_i,\theta_i), \gamma_n z_n^{-n+1} (r_i,\theta_i), \cdots, \gamma_n z_n^{+n} (r_i,\theta_i)]^T
\label{eq:pz_vec}
\end{align}
where $\gamma_n\defeq (n+1)/(\pi N)$ accounts for subsequent constants as well as integration to summation approximations in (\ref{eq:pz_mom}).
Note, we assume that $r_i\le 1$, $\forall i\in [1,N]$, which ensures that all image points are within the unit circle.
The PZ-polynomials based basis matrix can then be defined as a $P\times N$ matrix $\Zbf$, i.e., 
\be
\Zbf \defeq \left[\zbf_1, \zbf_2, \cdots, \zbf_N\right]
\label{eq:Z}
\ee
which is still a matrix of complex polynomials.
Now, given a set of training image measurements $\{\gbf_j^k\}$, for $j=1,\cdots,J_k$ and $k=1,\cdots,K$, the dictionary based on PZ-moments features, with the property of rotational invariance, can be defined as a column normalised (i.e., normalised to unity) $P\times J$ matrix $\Abf$, i.e., 
\be
\Abf \defeq {\rm abs} (\Zbf \Gbf)
= \left[ \Abf^1, \Abf^2, \cdots, \Abf^K \right]
\label{eq:pz_dict}
\ee
where ${\rm abs} (\cdot)$ is a function which generates element-wise absolute values, and $\Abf^k \defeq {\rm abs} (\Zbf \Gbf^k)$ is a $P\times J_k$ matrix of PZ-moments w.r.t. $\Gbf^k$, for $k=1,\cdots,K$.
In order to capitalise on the invariance structure provided by PZ-moments, we introduce auxiliary atoms in the dictionary (see Section \ref{sec:aux_atoms} for details). Thus, the dictionary can be defined as
\begin{align}
\Phibf &\defeq \left[ [\Abf^1, f(\Abf^1)], [\Abf^2, f(\Abf^2)], \cdots, [\Abf^K, f(\Abf^K)] \right] \notag \\
&= \left[ \Phibf ^1, \Phibf ^2, \cdots, \Phibf ^K \right]
\label{eq:pz_dict_mod}
\end{align}
where $f(\Abf^k)$ is a $P\times L_k$ auxiliary matrix (with columns normalised to unity) and it is a function of the columns of $\Abf^k$, $\Phibf ^k \defeq [\Abf^k, f(\Abf^k)]$ is a $P\times Q_k$ matrix with $Q_k = J_k + L_k$, for $k=1,\cdots,K$, and the over-complete dictionary $\Phibf$ is a $P\times Q$ matrix with $Q = \sum_{k=1}^K Q_k$. Now, the test image $\tilde{\ybf}$ can be encoded according to the following linear model.
\be
\ybf = \Phibf \xbf + \nbf
\label{eq:pz_model}
\ee
where $\ybf \defeq {\rm abs} (\Zbf \tilde{\ybf})$ is the $P\times 1$ vector of PZ-moments of the test image, $\xbf$ is the $Q\times 1$ encoded vector defined as,
$
\xbf \defeq [ \xbf^{1\,T}, \xbf^{2\,T}, \cdots, \xbf^{K\,T} ]^T
$,
where $\xbf^{k}$ is the $Q_k\times 1$ encoding vector w.r.t. $\Phibf ^k$, for $k=1,\cdots,K$, and $\nbf$ is the $P\times 1$ model error vector with bounded energy, i.e., $\| \nbf\|_2<\epsilon$. 
It is clear from (\ref{eq:pz_model}), given that the test image belongs to a particular class, $\xbf$ would be a sparse vector with nonzero elements ideally corresponding to the sub-dictionary of only a particular class.
\subsection{Sparse Reconstruction and Classification}
\label{sec:recon}

Since $P\ll Q$, (\ref{eq:pz_model}) is an under-determined system of linear equations. In order to recover $\xbf$ in (\ref{eq:pz_model}), we use IHT as the sparse recovery algorithm. An estimate of $\xbf$ can be obtained by processing the following iterations.
\be
\hat{\xbf}^{[t+1]} = \calH_\Gamma \left(\hat{\xbf}^{[t]} +  \Phibf^T \left(\ybf -  \Phibf \hat{\xbf}^{[t]}  \right) \right)
\label{eq:iht}
\ee
where $t$ is the iteration index (starting with $t=0$) and $\calH_\Gamma$ is the hard thresholding operator defined as
\be
\calH_\Gamma (\qbf) \defeq \qbf\setI_{\{ i \, | \, [\qbf]_i \ge [\text{Ascend}(\qbf)]_\Gamma , \, \forall i \}}
\ee
where $\setI_{\{\cdot\}}$ is an indicator operator which discards those elements of vector $\qbf$ that are not in the indicator set (the set given in its subscript), and $\text{Ascend}(\qbf)$ is a sorting function which sorts the elements of $\qbf$ in an ascending order. Essentially, $\calH_\Gamma (\qbf) $ preserves only the $\Gamma$ largest element magnitudes of $\qbf$ in each iteration $t$. Thus, (\ref{eq:iht}) approximates the $\ell_0$-norm estimate of $\xbf$, i.e., 
\be
\hat{\xbf} =  \arg \min_{\xbf}  \left\| \ybf - \Phibf \xbf \right\|_2^2 \quad \text{subject to} \;  \left\|\xbf\right\|_0^0 \le \Gamma
\label{eq:l0}
\ee
where $\Gamma$ is the order of sparsity. Note, the stopping criterion of iterations in (\ref{eq:iht}) can either be the maximum number of allowable iterations or the minimum residual error, i.e., $\|\ybf -  \Phibf \hat{\xbf}^{[t]}\|_2^2 / \|\ybf\|_2^2$.
After sparse encoding of $\ybf$, the classification of the target image can be done by solving the following OP.
\be
\hat{k} = \arg \min_{k}  \| \ybf - \Phibf^k \hat{\xbf}^k \|_2^2, \;\; \text{for\;\,} k=1, \cdots, K
\label{eq:classify_OP_2}
\ee
where $\hat{\xbf}^k$ is the estimate obtained in the PZ-SRC framework, when the stopping criterion for (\ref{eq:l0}) has been achieved.
\subsection{Auxiliary Atoms}
\label{sec:aux_atoms}
The auxiliary atoms can have a substantial impact on the performance of the classification. 
Ideally, variations in the image measurements w.r.t. different aspect angles, should not produce any variations in their respective PZ-moments. However, radar reflectivities at different aspect angles might not be uniform. Therefore, image at one aspect angle might be absolutely different from the image obtained at another aspect angle. Also, noise in the form of clutter or other artefacts can play a disruptive role. Auxiliary atoms try to recover the information lost due to these irregularities.
In this section, we present a number of techniques to generate the auxiliary atoms. Note, here our focus is primarily on rotational invariance of the moments.
\subsubsection{Fixed Auxiliary Atoms (${\rm AuxFix}$)}
\label{sec:aux_fix}

In case the measurements are obtained at random aspect angles, we propose to constitute the auxiliary atoms as an overall average of the PZ-moments based measurements of each class, i.e., 
\be
f(\Abf^k) = \sum_{j=1}^{J_k} \abf^k_j
\label{eq:aux_avg1}
\ee
where $\abf^k_j \defeq \text{abs} (\Zbf \gbf^k_j)$, for $k=1,\cdots,K$. ${\rm AuxFix}$ causes the effect of irregular reflectivities to be averaged out. 
Here, $L_k = 1$, for $k=1,\cdots,K$.
\subsubsection{Moving Average Based Auxiliary Atoms (${\rm AuxMov}$)}
\label{sec:aux_mov}

In case the measurements are arranged in the order of increasing aspect angles around the object, a moving average of atoms over each class can constitute the auxiliary atoms, i.e., 
\be
f_j(\Abf^k) \;\; = \sum_{w=-W_k/2}^{+W_k/2} \abf^k_{w+j}
\label{eq:aux_avg_mov}
\ee
where $W_k$ is the window size for the $k$th class, for $k=1,\cdots,K$, and $j=1,\cdots,J_k$. We can see that the window is centred over the $j$th column of $\Abf^k$. Note, in case $(w+j)<1$ or $(w+j)>J_k$, $\abf^k_{w+j}$ can be considered as zero vectors. 
Here, $L_k = J_k$. The auxiliary matrix can be formed as
\be
f(\Abf^k) = \left[  f_1(\Abf^k), f_2(\Abf^k), \cdots, f_{J_k}(\Abf^k) \right].
\label{eq:aux_mov}
\ee
\subsubsection{Correlation Based Auxiliary Atoms (${\rm AuxCorr}$)}
\label{sec:aux_corr}

An optimal method is to find correlated atoms w.r.t. every training measurement for each class, i.e., the columns of $\Abf^k$. The auxiliary atoms can then be generated based on a minimum correlation value, i.e., 
\be
f_j(\Abf^k) \;\; = \sum_{\begin{subarray}{1} \quad\;\, l=1 \\ \abf_j^{k\,T} \abf_l^k \, > \, \Upsilon \end{subarray}}^{J_k} \abf^k_l
\label{eq:aux_avg_corr}
\ee
where $\abf_j^{k\,T} \abf_l^k$ performs the inner product, $\Upsilon$ is the correlation threshold, and $j=1,\cdots,J_k$. 
Here, $L_k = J_k$. The auxiliary matrix can be formed according to (\ref{eq:aux_mov}).
This procedure ensures that all informative measurements, i.e., measurements with high mutual correlation, are accounted for.
\subsection{Complex Signatures}
\label{sec:compl}

We can see from the previous sections that most of the classification strategies use only the intensities or magnitudes of the images. However, a radar signature contains information both in the magnitude as well as in the phase. To this end, we combine the magnitude and the phase of the radar signatures via an averaging fusion metric, and use the fused image to create the PZ-moments. Thus, the fused image has the form, $0.5[{\rm abs} (\{\gbf_j^k\}) + {\rm phase} (\{\gbf_j^k\})] $, where ${\rm phase} (\cdot)$ is an element-wise phase generating function.
\section{Simulations}
\label{sec:sim}

\begin{figure}[t]
\centering
\begin{subfigure}[b]{0.3\linewidth}
    \centering
     \includegraphics[width=.99\textwidth]{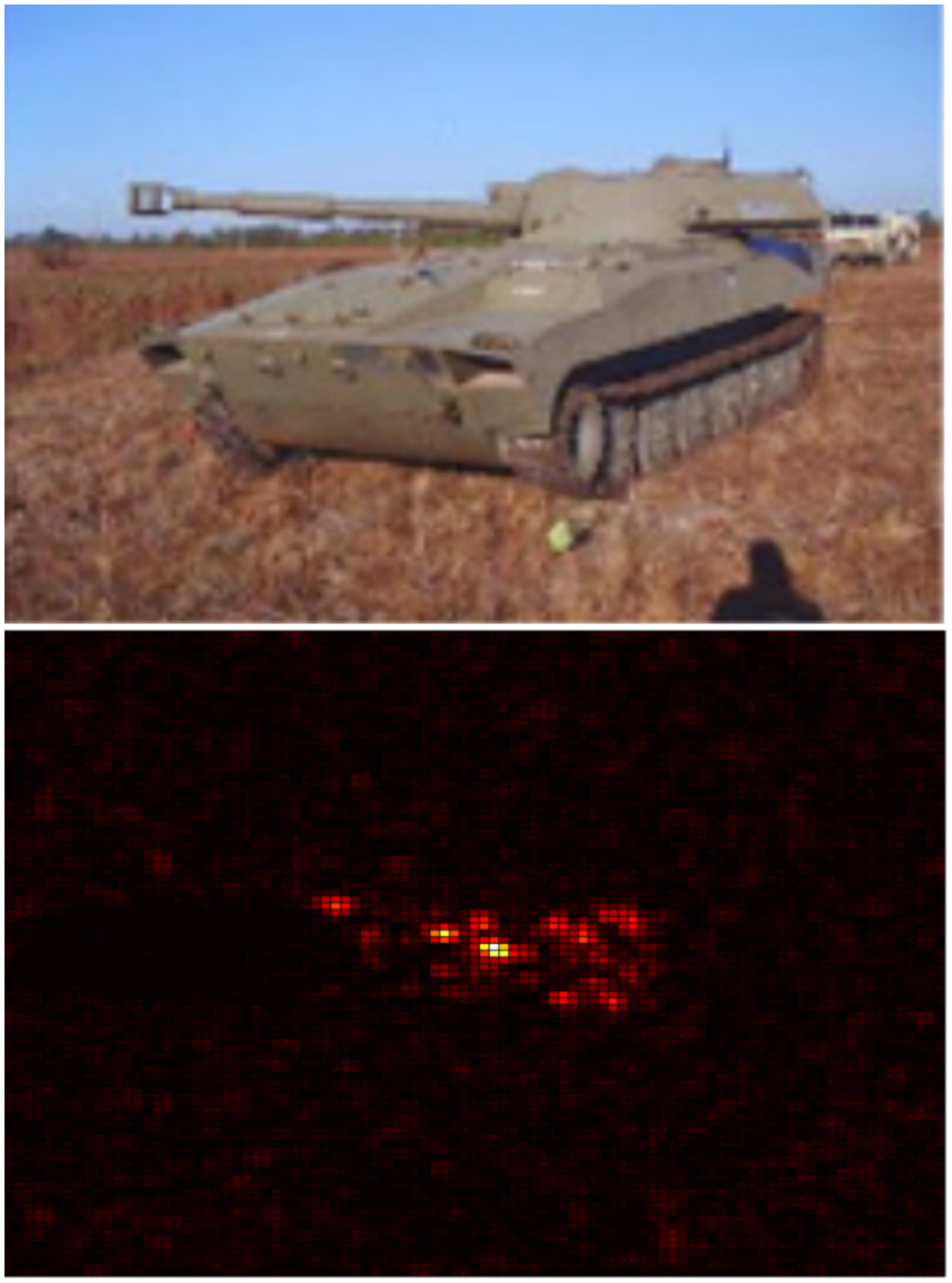} 
    \caption{2S1}\label{fig:t00}
  \end{subfigure}
  \begin{subfigure}[b]{0.3\linewidth}
    \centering
     \includegraphics[width=.99\textwidth]{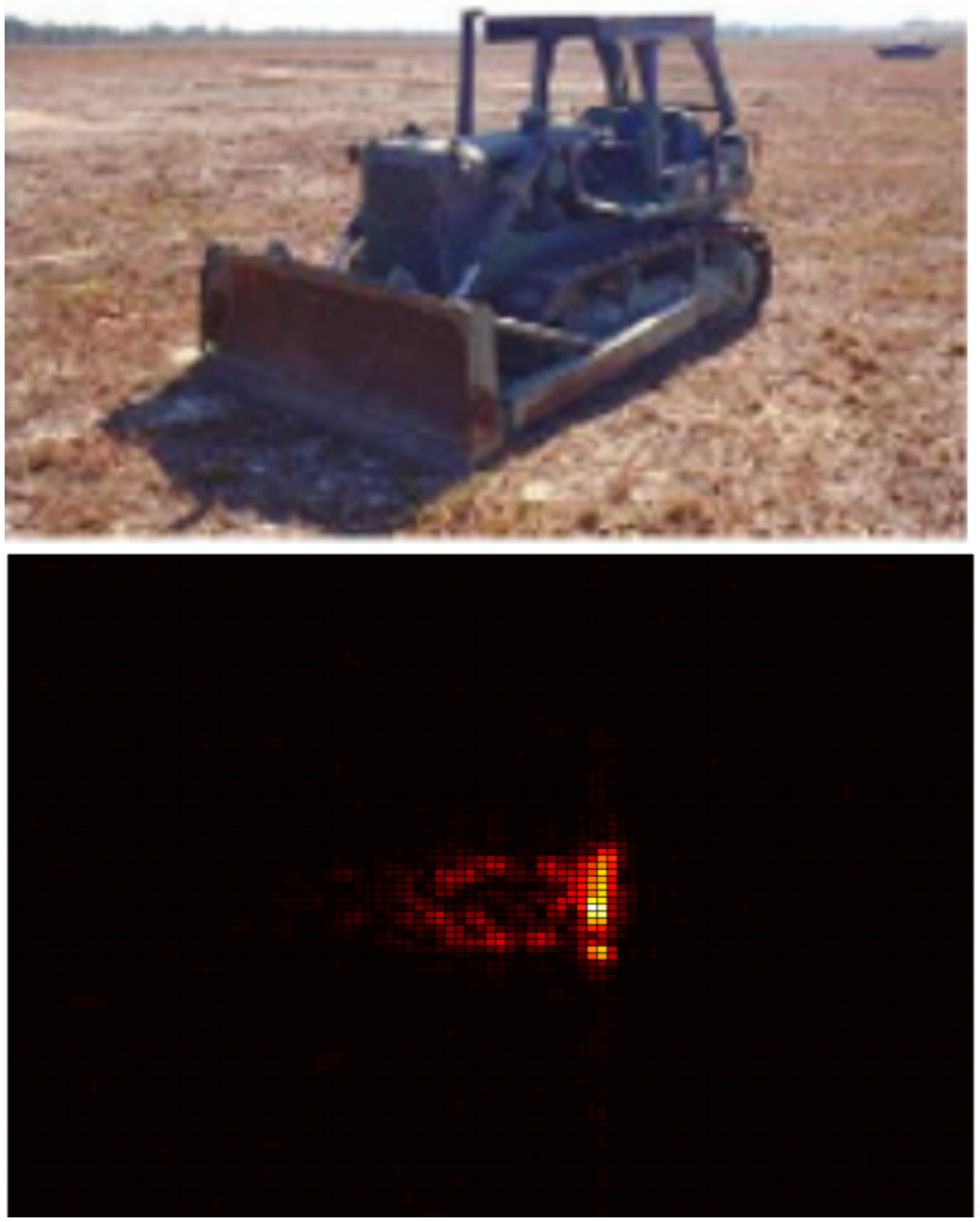} 
    \caption{D-7}\label{fig:t05}
  \end{subfigure}
  \begin{subfigure}[b]{0.3\linewidth}
    \centering
     \includegraphics[width=.99\textwidth]{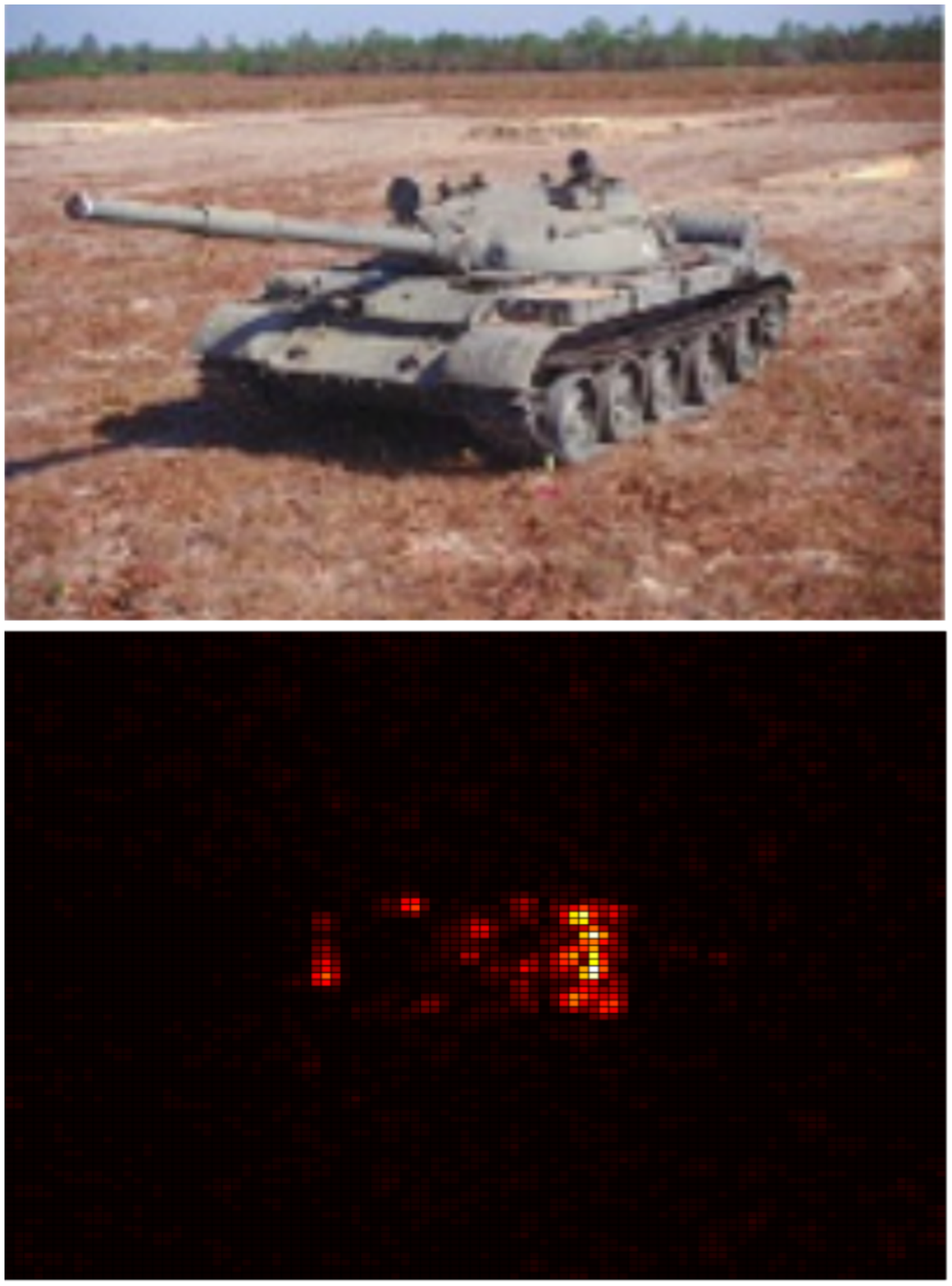} 
    \caption{T62}\label{fig:t16}
  \end{subfigure}
  \caption{MSTAR Targets.}
  \label{fig:targets}
\end{figure}

\begin{figure}[tb]
\centering
\includegraphics[width=.99\linewidth]{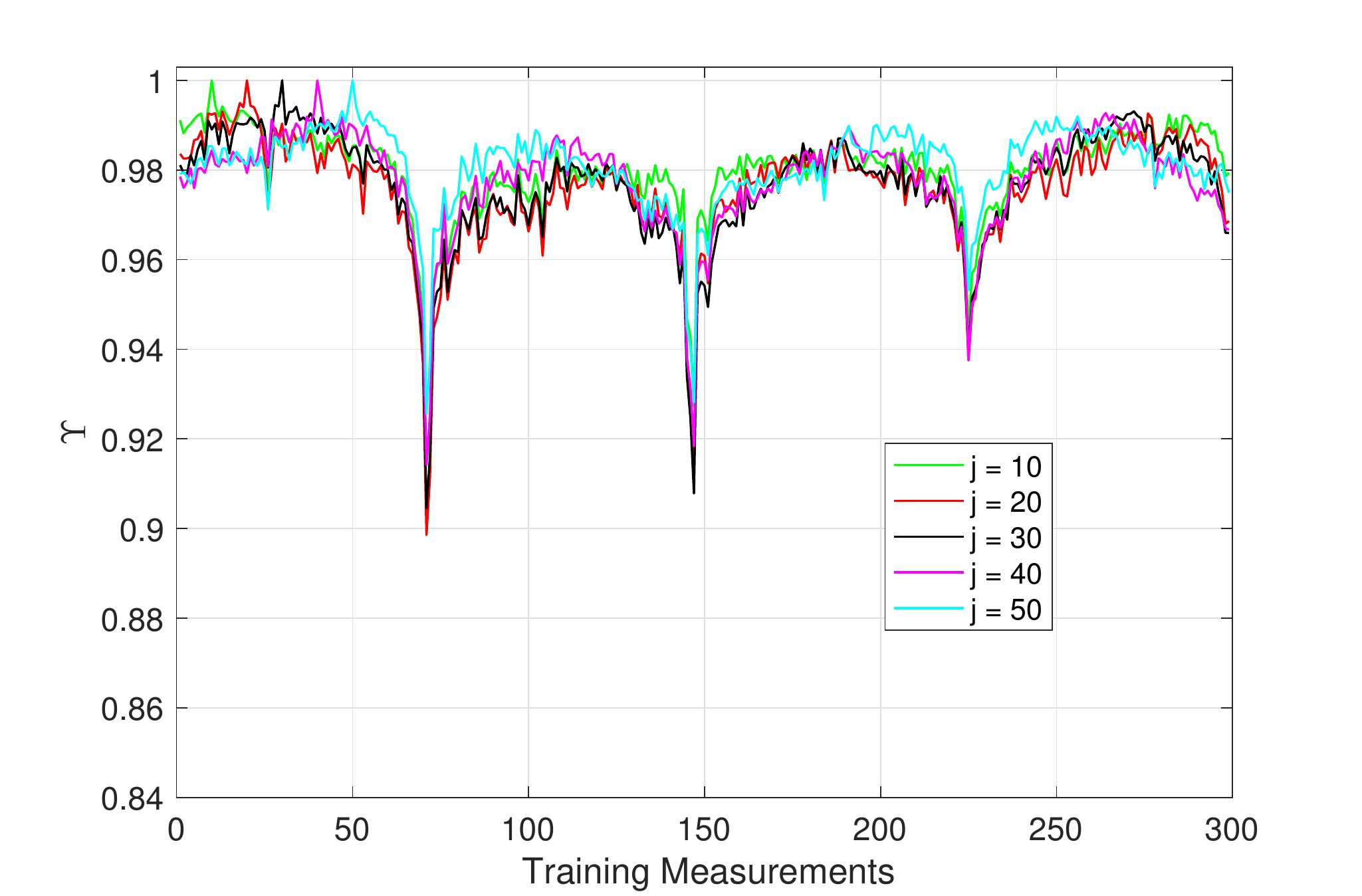} 
\caption{Correlations among PZ-moments based Measurements.}
\label{fig:corr_pz_dict}
\end{figure}

In this section we present simulation results of our proposed methods. 
We use the publicly available MSTAR dataset. 
We consider three targets, i.e., 2S1 tank, D-7 land clearing vehicle and T62 tank (so $K=3$). Figure \ref{fig:targets} shows the optical and SAR (magnitude only) images, for one aspect angle of these targets. 
For the purpose of training, a total of $J_k=299$ measurements are considered, for each target, at a radar elevation angle of $17^\circ$. The measurements have been taken at sequentially increasing aspect angles of approximately $1.2^\circ$, i.e, covering the complete angular range of $360^\circ$. 
Note, the measurements are in the form of $96\times 96$ SAR images. These images are vectorised for the sake of processing. Thus, $N=9216$.
For the purpose of testing, a total of $273$ image measurements (for each class) are considered, which have been taken at different aspect angles over the complete angular range of $360^\circ$, with a radar elevation angle of $15^\circ$. Aspect angles of the testing measurements are different from the training measurements. Thus, pose angle estimation is a valid issue. 
We define the classification/recognition accuracy/performance for the $k$th class as
\be
\Omega_k \defeq 100\left(\dfrac{{\rm TP}_k}{273}\right)
\label{eq:omeg_k}
\ee
where ${\rm TP}_k$ are the true positives of the target class $k$, for $k=1,\cdots,K$, and the overall performance is defined as
\be
\Omega \defeq \dfrac{1}{K} \sum_{k=1}^K \Omega_k.
\label{eq:omeg}
\ee
Note, both $\Omega$ and $\Omega_k$ quantify performance in percentages. For PZ-moments, we consider $n=10$ as the degree of the polynomials which generates $P=121$ PZ-moments. In comparison to $N$, this is a dimensionality reduction of over $98\%$.
Note, the value of $n$ can impact the performance of classification. Generally, higher values of $n$ can represent an image better. However, very large values can cause numerical instabilities. Therefore, we select a moderate value of $n$. Few tests on the training data can also give a good idea over the choice of $n$.
For sparse reconstruction, we use IHT for PZ-SRC (as well as for SRC, for a fair comparison) and consider the order of sparsity $\Gamma=5$. 
Note, the parameter $\Gamma$ is a tuning parameter and can be selected based on different cross validation approaches.
\\
\begin{table}
\renewcommand{\arraystretch}{1.3}
\begin{center}
\caption{Performance Comparison of Different Classifiers}
\begin{tabular}{ c | c | c | c | c}
\hline
{} & $\Omega_1 (\%)$ & $\Omega_2(\%)$ & $\Omega_3(\%)$ & $\Omega(\%)$ \\ \hline
LSVC & $91.94$ & $98.90$ & $92.67$  & $94.50$ \\ \hline
SRC & $95.97$ & $99.26$ & $93.04$  & $96.09$ \\ \hline
PZ-LSVC & $93.40$ & $99.26$ & $96.33$ & $96.33$ \\ \hline
PZ-SRC& $96.70$ & $99.26$ & $96.33$  & $97.43$ \\ \hline
\end{tabular}
\label{tab:compare_classify}
\end{center}
\end{table}
\begin{table}
\renewcommand{\arraystretch}{1.3}
\begin{center}
\caption{Confusion Matrix for PZ-SRC (Magnitude Only)}
\begin{tabular}{ c | c | c | c | c | c}
\hline
{} & 2S1 & D-7 & T62 & $\Omega_k(\%)$ & $\Omega(\%)$ \\ \hline
2S1 & $264$ & $0$ & $9$  & $96.70$ & {-} \\ \hline
D-7 & $0$ & $271$ & $2$  & $99.26$ & {-} \\ \hline
T62 & $2$ & $8$ & $263$  & $96.33$ & {-} \\ \hline
 {} & {-} & {-} & {-}  & {-} & $97.43$ \\ \hline
\end{tabular}
\label{tab:confMtx_pzsrc_mag}
\end{center}
\end{table}
\begin{table}
\renewcommand{\arraystretch}{1.3}
\begin{center}
\caption{Confusion Matrix for PZ-SRC (Complex)}
\begin{tabular}{ c | c | c | c | c |c }
\hline
{} & 2S1 & D-7 & T62 & $\Omega_k(\%)$ & $\Omega(\%)$ \\ \hline
2S1 & $263$ & $0$ & $10$  & $96.33$ & {-} \\ \hline
D-7 & $0$ & $271$ & $2$  & $99.26$ & {-} \\ \hline
T62 & $0$ & $1$ & $272$  & $99.63$ & {-} \\ \hline
 {} & {-} & {-} & {-}  & {-} & $98.41$ \\ \hline
\end{tabular}
\label{tab:confMtx_pzsrc_complx}
\end{center}
\end{table}
In terms of simulation results, we first consider the magnitude only radar signatures. Table \ref{tab:compare_classify} shows the classification performance results of different classifiers. Note, we also consider PZ-moments based LSVC (PZ-LSVC) for the sake of comparison.
We can see that SRC outperforms LSVC for all target classes. PZ-LSVC is slightly better than SRC in the overall performance. However, PZ-SRC shows better classification performance than all the classifiers, in every category. In general, PZ-moments based methods have an edge over other methods, despite the dimensionality reduction.
Table \ref{tab:confMtx_pzsrc_mag} provides the confusion matrix for PZ-SRC (magnitude only). Performance of each target is given in their respective rows. Each column shows the number of test images classified as the title target. Last two columns show the classification performance of individual targets and the overall, respectively.
Table \ref{tab:confMtx_pzsrc_complx} provides the confusion matrix for complex radar signatures, where we use the fusion technique of Section \ref{sec:compl}. We see an improved overall performance of $98.41\%$ in comparison to $97.43\%$ of the magnitude only in Table \ref{tab:confMtx_pzsrc_mag}.
\\
For the rest of the simulations, we use fused complex signatures. We first obtain classification results by considering ${\rm AuxFix}$ of Section \ref{sec:aux_fix} as auxiliary atoms. Table \ref{tab:confMtx_pzsrc_AuxFix} shows the confusion matrix in this regard.
The performance improvement has been encouraging, with $\Omega=98.53\%$. 
Next, we simulate the classification problem by considering ${\rm AuxMov}$ of Section \ref{sec:aux_mov} as auxiliary atoms. Table \ref{tab:vary_wind} shows the performance of PZ-SRC for varying sizes of $W_k$ (same $\forall k$). We can see that the classification performance is affected by changing size of $W_k$. The best performance is achieved when $W_k/J_k$ is a multiple of $0.5$.
If all the test measurements are divided into four quadrants, with each quadrant corresponding to a range of aspect angles of approximately $90^\circ$, then $W_k/J_k=0.5$ essentially corresponds to the numerical size of one quadrant, when the best performance is achieved.
\begin{table}
\renewcommand{\arraystretch}{1.3}
\begin{center}
\caption{Confusion Matrix for PZ-SRC (${\rm AuxFix}$)}
\begin{tabular}{ c | c | c | c | c | c}
\hline
{} & 2S1 & D-7 & T62 & $\Omega_k(\%)$ & $\Omega(\%)$\\ \hline
2S1 & $264$ & $0$ & $9$  & $96.70$ & {-} \\ \hline
D-7 & $0$ & $271$ & $2$  & $99.26$ & {-} \\ \hline
T62 & $0$ & $1$ & $272$  & $99.63$ & {-} \\ \hline
 {} & {-} & {-} & {-}  & {-} & $98.53$ \\ \hline
\end{tabular}
\label{tab:confMtx_pzsrc_AuxFix}
\end{center}
\end{table}
\begin{table*}
\renewcommand{\arraystretch}{1.3}
\begin{center}
\caption{Performance with Varying $W_k$ in (\ref{eq:aux_avg_mov})}
\begin{tabular}{ c | c | c | c | c | c | c | c | c | c | c | c | c | c | c | c}
\hline
$W_k$  & $29$ & $59$ & $89$  & $119$ & $149$ & $179$ & $209$ & $239$ & $269$ & $299$ & $328$ & $358$  & $388$ & $418$ & $448$\\ \hline
${W_k}/{J_k}$ & $0.1$  & $0.2$  & $0.3$ & $0.4$ & $0.5$ & $0.6$ & $0.7$ & $0.8$ & $0.9$ & $1.0$ & $1.1$ & $1.2$ & $1.3$ & $1.4$ & $1.5$\\ \hline
$\Omega(\%)$ & $98.16$ & $98.41$ & $98.65$ & $98.77$  & $98.90$ & $98.65$ & $98.65$ & $98.77$ & $98.90$ & $98.90$  & $98.90$ & $98.77$ & $98.65$ & $98.77$ & $98.90$\\ \hline
\end{tabular}
\label{tab:vary_wind}
\end{center}
\end{table*}
This can be explained as follows. Due to the rotational invariance properties of the PZ-moments, measurements at consecutive aspect angles are correlated to each other, in general, with some variations mostly because of radar reflectivity irregularities. However, measurements at the boundary of two quadrants correspond to the fine corners of the considered rectangular-shaped targets, and these measurements are highly uncorrelated with all the measurements in the preceding or the succeeding quadrant. This phenomenon can be seen in Figure \ref{fig:corr_pz_dict}, which shows the correlations of a few training measurements (PZ-moments) with the rest of the measurements in a $k$th target class. We can see that mutual correlations are minimum at quadrants, i.e., when $W_k/2 = 75, 150, 225$.
Thus, it is better to exploit only the correlated measurements for generating auxiliary atoms and that happens when the size of $W_k$ is such that it contains most of the correlated measurements of a quadrant or its multiple.
\begin{figure}[tb]
\centering
\includegraphics[width=.99\linewidth]{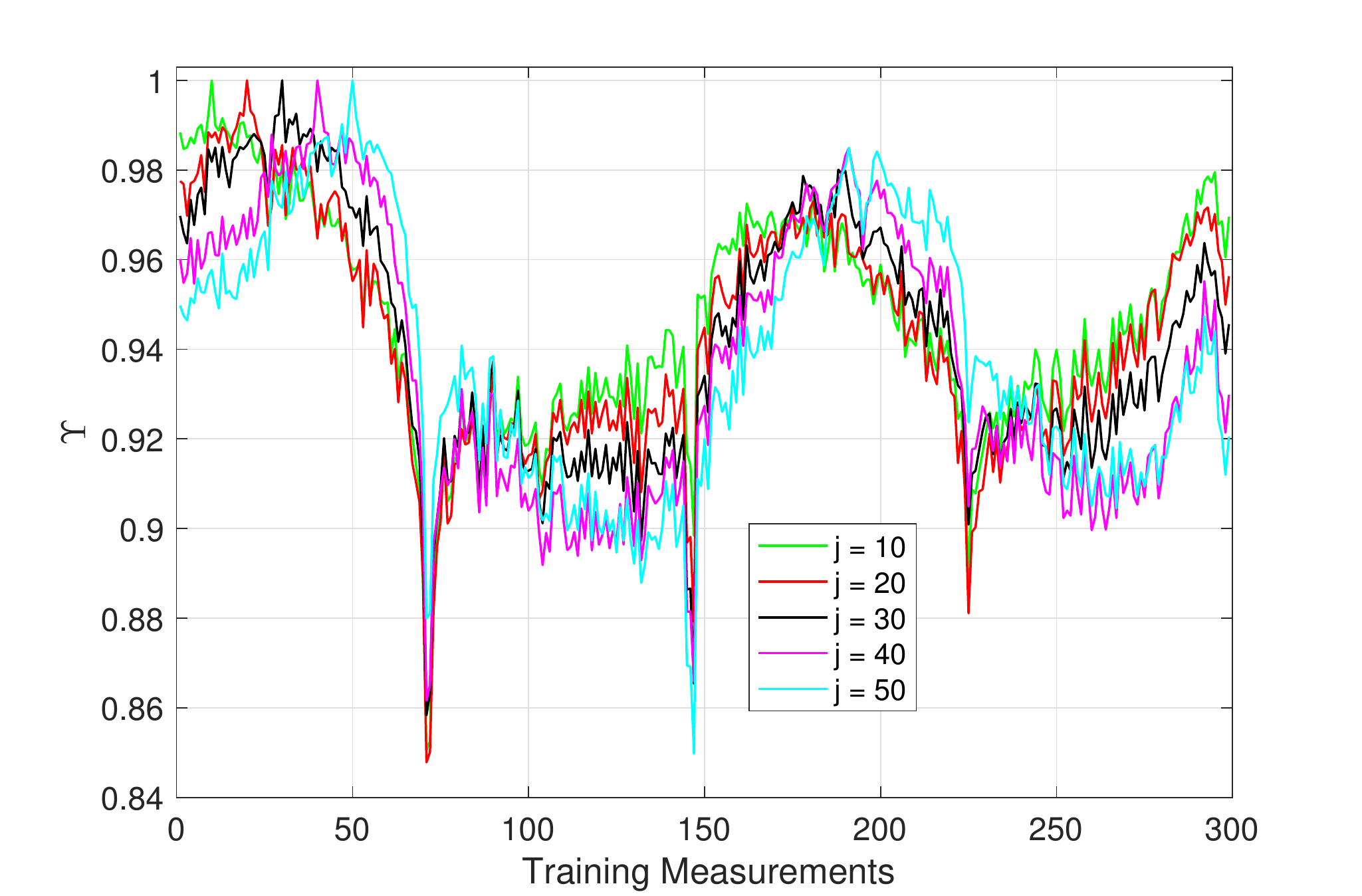} 
\caption{Correlations among Test Measurements.}
\label{fig:corr_dict}
\end{figure}
Since, correlation is an important parameter for generating auxiliary atoms, we next assess the classification performance by considering ${\rm AuxCorr}$ of Section \ref{sec:aux_corr}. Table \ref{tab:vary_corr} shows the classification performance with varying $\Upsilon$. We can see that best performance is achieved for $\Upsilon=0.94$. This is quite understandable. A higher value of $\Upsilon$ does not collect enough number of informative measurements and a lower value of $\Upsilon$ involves noisy or non-informative measurements. This can be seen from Figure \ref{fig:corr_pz_dict} as well. 
We also provide a confusion matrix regarding the performance of PZ-SRC with fused complex signatures and using $W_k/J_k = 0.5$, in Table \ref{tab:confMtx_pzsrc}. An overall performance of $98.90\%$ is achieved.
Note, in order to better appreciate the invariance properties of the PZ-moments, we also plot the correlations among original test measurements, i.e., without PZ-moments, in Figure \ref{fig:corr_dict}. We can see that the correlation structure is quite inconsistent in comparison to the PZ-moments as in Figure \ref{fig:corr_pz_dict}.  
\begin{table}
\renewcommand{\arraystretch}{1.3}
\begin{center}
\caption{Performance with Varying $\Upsilon$ in (\ref{eq:aux_avg_corr})}
\begin{tabular}{ c | c | c | c | c | c | c | c  }
\hline
$\Upsilon$ & $0.98$ & $0.97$ & $0.96$ & $0.95$ & $0.94$ & $0.93$ & $0.92$\\ \hline
$\Omega(\%)$ & $98.29$ & $98.53$ & $98.65$  & $98.77$ & $98.90$ & $98.53$ & $98.53$\\ \hline
\end{tabular}
\label{tab:vary_corr}
\end{center}
\end{table}
\begin{table}
\renewcommand{\arraystretch}{1.3}
\begin{center}
\caption{Confusion Matrix for PZ-SRC (${\rm AuxMov}$)}
\begin{tabular}{ c | c | c | c | c | c}
\hline
{} & 2S1 & D-7 & T62 & $\Omega_k(\%)$ & $\Omega(\%)$ \\ \hline
2S1 & $266$ & $0$ & $7$  & $97.43$ & {-} \\ \hline
D-7 & $0$ & $271$ & $2$  & $99.26$ & {-} \\ \hline
T62 & $0$ & $0$ & $273$  & $100$ & {-} \\ \hline
 {} & {-} & {-} & {-}  & {-} & $98.90$  \\ \hline
\end{tabular}
\label{tab:confMtx_pzsrc}
\end{center}
\end{table}
\section{Conclusions}
\label{sec:concl}
In this paper, we presented sparse representations for radar image classification by using pseudo-Zernike moments. We obtained a reduction in the dimensionality of the problem without compromising the performance. We exploited the invariance properties of the pseudo-Zernike moments to generate auxiliary atoms to complement the dictionary, which resulted in an enhanced classification performance. We used a fusion strategy to gain both from the magnitude as well as the phase of the radar signatures. We proved the validity of our proposed methods via simulations on the MSTAR dataset.
\section*{Acknowledgements}
This work has been approved for submission by TASSC-PATHCAD Sponsor, Chris Holmes, Senior Manager Research, Research Department, Jaguar Land Rover, Coventry, UK.
\bibliographystyle{IEEEtran}

\end{document}